 \documentclass[aps,pre,onecolumn,superscriptaddress,nofootinbib,10pt,a4]{revtex4}

\usepackage{bm}
\usepackage{amsfonts}
\usepackage{graphicx}        % standard LaTeX graphics tool
\newcommand{\rme}{{\mathrm{e}}}

\newcommand{\rmpp}{{\mathrm{p}}}
\newcommand{\rmd}{{\mathrm{d}}}
 \newcommand{\rmi}{{\mathrm{i}}}
\newcommand{\rmD}{{\mathrm{D}}}

\newcommand{\iint}{{\int\!\!\!\int}}   % deja dans RevTeX

\newcommand{\bfa}{\mathbf{a}}

\newcommand{\bfk}{\mathbf{k}}
\newcommand{\bfl}{\mathbf{l}}
\newcommand{\bfm}{\mathbf{m}}
\newcommand{\bfn}{\mathbf{n}}
\newcommand{\bfp}{\mathbf{p}}
\newcommand{\bfr}{\mathbf{r}}

\newcommand{\bfv}{\mathbf{v}}

\begin{document}

\title[Direct path from microscopic mechanics to Debye shielding and Landau damping]{Direct
path from microscopic mechanics to Debye shielding, Landau damping,
and wave-particle interaction}

\author{D~F Escande, Yves Elskens and F Doveil}

%%% \affiliation{Aix-Marseille Universit\'{e}, CNRS, PIIM, UMR 7345, \\

\affiliation{Aix-Marseille Universit\'{e}, CNRS, PIIM, UMR 7345, \\
   case 321, campus Saint-J\'er\^ome, FR-13013 Marseille, France}
%%% \email{Dominique.Escande@univ-amu.fr}
\email{{dominique.escande@univ-amu.fr},
         {fabrice.doveil@univ-amu.fr},
          {yves.elskens@univ-amu.fr}}

\begin{abstract}
The derivation of Debye shielding and Landau damping from the $N$-body description of plasmas is performed directly
by using Newton's second law for the $N$-body system.
This is done in a few steps with elementary calculations using standard tools of calculus, and no probabilistic setting.
Unexpectedly, Debye shielding is encountered together with Landau damping.
This approach is shown to be justified in the one-dimensional case when the number of particles in a Debye sphere becomes large.
The theory is extended to accommodate a correct description of trapping and chaos due to Langmuir waves.
Shielding and collisional transport are found to be two related aspects of the repulsive deflections of electrons,
in such a way that each particle is shielded by all other ones while keeping in uninterrupted motion.
%[\today]

\par\medskip
\noindent
  PACS numbers :  \newline %
  52.20.-j   Elementary processes in plasmas \newline
  52.35.Fp  Plasma : electrostatic waves and oscillations \newline
  45.50.-j   Dynamics and kinematics of a particle and a system of particles \newline
  05.20.Dd  Kinetic theory \newline

\par\medskip
\noindent
{\textit{Keywords}} :
  basic plasma physics,
  Debye shielding,
  Landau damping,
  wave-particle interaction,
  spontaneous emission,
  amplitude equation,
  N-body dynamics

\end{abstract}

\par \medskip
%  \noindent  %\mathbf{preprint submitted for publication} \\
%   \textbf{DRAFT -- private communication, not for distribution}
%  % To appear in: \textit{***}.
%  \par

\maketitle

%VVVVVVVVVVVVVVVVVVVVVVVVVVVVVVVVVVVVVVV
\section{Introduction}
\label{secIntro}
%VVVVVVVVVVVVVVVVVVVVVVVVVVVVVVVVVVVVVVV

For macroscopic classical systems, the $N$-body description by classical mechanics was deemed impossible. This led to the development of thermodynamics, of fluid mechanics, and of kinetic equations to describe various macroscopic systems made up of particles like electrons, gas atoms or molecules, stars, or microorganisms. When plasma physicists had to address the microscopic description of their state(s) of matter, they did not consider the $N$-body description by classical mechanics, but directly derived kinetic analogues of the Boltzmann equation, in particular the Vlasov equation. This trend has been dominant till nowadays.

However, for plasmas where transport due to short range interactions is weak, $N$-body classical mechanics yields useful results.
As will be recalled in section \ref{WPDRT}, it already enabled a description of wave-particle interaction making it more intuitive, incorporating modern chaotic dynamics, and unifying particle and wave evolutions, as well as collective and finite-$N$  physics \cite{AEE,EZE,EEbook,Houches}.
The present paper makes an even more thorough use of $N$-body mechanics by working directly with Newton's second law for this system. It shows, in particular, that basic phenomena like Debye shielding and Landau damping can be easily derived while avoiding kinetic and statistical calculations altogether.
Furthermore, the new derivation brings together Debye shielding and Landau damping, a totally unexpected fact in view of presentations in classical textbooks.

%VVVVVVVVVVVVVVVVVVVVVVVVVVVVVVVVVVVVVVV
\section{Main results and paper outline}
\label{MR}
%VVVVVVVVVVVVVVVVVVVVVVVVVVVVVVVVVVVVVVV

Here are the main results of this paper and its organization~:
\begin{enumerate}
\item{\label{claim-1}
Section \ref{FEP} analyses a perturbation from ballistic orbits of an infinite plasma made up of the periodic replication of $N$ electrons
coupled by Coulomb forces in a cubic volume $L^3$ with a neutralizing ionic background
(One Component Plasma (OCP) model \cite{Salp,Abe,BH,Kie14}).
By using the Fourier and Laplace transforms in a way similar to that in the Vlasovian derivation of Landau damping,
an equation (Eq.~(\ref{phihat})) is derived for a version of the electrostatic potential linearized from ballistic orbits.
This equation has the form ${\mathcal{E}} \phi= {\mathcal{S}}$, where ${\mathcal{E}}$ is a linear operator,
acting on the infinite dimensional array $\phi$ whose components are all the Doppler shifted Fourier-Laplace components of the potential.
Both ${\mathcal{E}}$ and the source term ${\mathcal{S}}$ are sums over the $N$ particles.
}
\item{\label{claim-2}
In section \ref{SCP}, the discrete sums in ${\mathcal{E}}$  are substituted with integrals over a continuous uniform distribution function $f_0(\bfv)$.
Then ${\mathcal{E}}$ becomes diagonal, which brings Eq.\ (\ref{phihatL}), $\epsilon(\bfm,\omega) \, \Phi(\bfm,\omega)
  = \phi^{(\rm{bal})}(\bfm,\omega)$, where $\epsilon(\bfm,\omega)$ is the classical plasma dielectric function, familiar in a Vlasovian context.
  Then, the new approximate potential $\Phi(\bfm,t)$ turns out to be the sum of two parts.
The first one is the sum of the shielded Coulomb potentials of the individual particles (Eq.~(\ref{phi})).
Such potentials were first computed by a kinetic approach in section II.A of Ref.\ \cite{Gasio} and later on in \cite{Bal,Rost}.
So, Debye shielding is computed for a single mechanical realization of the plasma.
}
\item{\label{claim-3}
Section \ref{STSP} describes the second part of $\Phi(\bfm,t)$.
It is the sum of the contributions to the excitation of Langmuir waves by the individual particles (Eq.\ (\ref{Phijt})).
Substituting again the discrete sums over particles with integrals over a continuous distribution function $f(\bfr,\bfv)$
yields the total amplitude of this wave in a compact way (Eq.\ (\ref{Phift})),
which works also for distributions that are non-smooth in $\bfv$ (for instance a two-stream one).
}
\item{\label{claim-3a}
In order to recover the usual expressions of textbooks,
section \ref{Vfor} introduces $f(\bfr,\bfv)$ before performing the inverse Laplace transform providing $\Phi(\bfm,t)$.
This yields Eqs (\ref{phihatL}) and (\ref{phi0hatcg}) which are the expressions including initial conditions in Landau contour calculations of Langmuir wave growth or damping,
obtained from linearizing Vlasov equation and using Fourier-Laplace transform,
as described in many textbooks (see for instance Refs \cite{HW,Nicholson,BoydSan}).
}
\item{\label{claim-3b}
Section \ref{Mrdrv} shows that the previous analysis can be performed for a smoothed Coulomb potential
whose Fourier expansion is bounded to wavenumbers smaller than $1/b_{\mathrm{smooth}}$,
where $b_{\mathrm{smooth}} \ll \lambda_\rmD$. Then the previous linearization is justified.
An argument developed in \ref{DSmoo} shows that the presence of many particles in the Debye sphere
justifies the passage to a continuous distribution function in the one-dimensional case.
}
\item{\label{claim-4}
In section \ref{MIIDS}, Picard iteration technique (one of the standard methods
to prove the existence and uniqueness of solutions to first-order equations with given initial conditions)
is applied to the equation of motion of a particle $P$ due to the Coulomb forces of all other ones.
It stresses that a part of the effect on particle $P$ of another particle $P'$
is mediated by all other particles (Eq.~(\ref{rsecnAccDev2})) and reduces the direct part.
Indeed, particle $P'$ modifies the motion of all other particles,
implying that the action of the latter ones on particle $P$ is affected by particle $P'$.
}
\item{\label{claim-5}
This calculation yields the following interpretation of shielding.
At $t = 0$, consider a set of uniformly distributed particles, and especially particle $P$.
At a later time $t$, the latter has deflected all particles which made a closest approach to it
with a typical impact parameter $b \lesssim v_{\rm{th}} t$, where $v_{\rm{th}}$ is the thermal velocity.
This \emph{part of their global deflection} due to particle $P$ reduces the number of particles
inside the sphere $S_P(t)$ of radius $v_{\rm{th}} t$ about it.
Therefore, according to Gauss' theorem, the effective charge of particle $P$ as seen out of $S_P(t)$ is reduced~:
the charge of particle $P$ is shielded due to these deflections.
This shielding effect increases with $t$, and thus with the distance to particle $P$.
It becomes complete at a distance on the order of $\lambda_\rmD$.
Since the global deflection of particles includes the contributions of many other ones, the density of the electrons does not change,
at variance with the shielding at work next to a probe (see e.g. section 2.2.1 of \cite{APiel}).

When starting from random particle positions, the typical time-scale for shielding to set in is the time for a thermal particle to cross a Debye sphere, i.e.\ $\omega_{\rmpp}^{-1}$,  where $\omega_{\rmpp}$ is the plasma frequency.
Furthermore, shielding, though very fast a process, is a cooperative dynamical one, not a collective (viz.\ coherent) one~:
it results from the accumulation of almost independent repulsive deflections with the same qualitative impact on the effective electric field of particle $P$ (if point-like ions were present, the attractive deflection of charges with opposite signs would have the same effect).
So, shielding and collisional transport are two aspects of the same two-body repulsive process.
}
\item{\label{claim-6}
In section \ref{WPDRT}, in the spirit of Refs \cite{OWM,OLMSS,AEE,EEbook}, to accommodate a correct description of trapping or chaos due to Langmuir waves, the set of particles is split into bulk and tail, where the bulk is the set of particles which cannot resonate with Langmuir waves.
Repeating for the bulk particles the analysis leading to Eq.~(\ref{phihat}), the same equation is recovered with an additional source term due to the tail particles (Eq.~(\ref{phihatU})).
}
\item{\label{claim-7}
Using the fact that the number of tail particles is small with respect to the bulk one, and a technique introduced in Refs \cite{OWM,OLMSS}, an amplitude equation is derived for any Fourier component of the potential where tail particles provide a source term (Eq.~(\ref{eqampl})).
}
\item{\label{claim-8}
This equation, together with the equation of motion of the tail particles, enables one to show that, in the linear regime, the amplitude of a Langmuir wave is ruled by Landau growth or damping, and by spontaneous emission (Eq.~(\ref{evampfinal})), a generalization to 3 dimensions of the one-dimensional result of Refs \cite{EZE,EEbook}.
}

\end{enumerate}

%VVVVVVVVVVVVVVVVVVVVVVVVVVVVVVVVVVVVVVV
\section{Fundamental linear equation for the potential}
\label{FEP}
%VVVVVVVVVVVVVVVVVVVVVVVVVVVVVVVVVVVVVVV

This paper deals with the One Component Plasma (OCP) model \cite{Salp,Abe,BH}, which considers the plasma as infinite with spatial periodicity $L$ in three orthogonal directions with coordinates $(x,y,z)$, and made up of $N$ electrons in each elementary cube with volume $L^3$. Ions are present only as a uniform neutralizing background, enabling periodic boundary conditions.
This choice is made to simplify the analysis which focuses on $\varphi(\bfr)$, the potential created by the $N$ particles at any point where there is no particle.
The discrete Fourier transform of $\varphi$, readily obtained from the Poisson equation, is given by $\tilde{\varphi}(\mathbf{0}) = 0$, and for $\bfm \neq \mathbf{0}$ by
\begin{equation}
  \tilde{\varphi}(\bfm)
  = -\frac{e}{\epsilon_0 k_{\bfm}^2} \sum_{j \in S}
     \exp(- \rmi \bfk_{\bfm} \cdot \bfr_j),
\label{phitildetotM}
\end{equation}
where $-e$ is the electron charge, $\epsilon_0$ is the vacuum permittivity, $\bfr_j$ is the position of particle $j$,
$S = \{ 1, \ldots N \}$, $\tilde{\varphi}(\bfm)
= \int \varphi(\bfr) \exp(- \rmi \bfk_{\bfm} \cdot \bfr) \, \rmd^3 \bfr$,
with $\bfm = (m_x,m_y,m_z)$ a vector with three integer components running from $- \infty$ to $+ \infty$, $\bfk_{\bfm} = \frac{2 \pi}{L} \, \bfm$, and $k_{\bfm} = \|\bfk_{\bfm}\|$. Reciprocally,
\begin{equation}
\varphi(\bfr) = \frac{1}{L^3}\sum_{\bfm} \tilde{\varphi} (\bfm) \exp(\rmi \bfk_{\bfm} \cdot \bfr) .
\label{phiInv}
\end{equation}

The dynamics of particle $l$ follows Newton's equation
\begin{equation}
  \ddot{\bfr}_l
  = \frac{e}{m_\rme} \nabla \varphi_l(\bfr_l),
\label{rsectot}
\end{equation}
with $m_\rme$ the electron mass, and $\varphi_l$ the electrostatic potential acting on particle $l$, i.e.\ the one created by all other particles and by
the background charge.
Its Fourier transform is given by Eq.\ (\ref{phitildetotM})
with the restriction $j \neq l$ in order to exclude self-interaction.
Let
\begin{equation}
  \bfr_l^{(0)}
  = \bfr_{l0} + \bfv_{l} t
\label{rl0}
\end{equation}
be a ballistic approximation to the motion of particle $l$, and let $\bm{\xi}_l = \bfr_l - \bfr_l^{(0)}$.
We leave some freedom to this approximation~: $\bfr_{l0}$ and $\bfv_{l}$ may be respectively the initial position and velocity of particle $l$,
or they may be slightly shifted from these values, for instance by low amplitude Langmuir waves.
We now define the ballistic approximation $\phi_l^{(\rm{bal})} (\bfm,t)$
to $\tilde{\varphi}_l (\bfm,t)$ which is computed from Eq.\ (\ref{phitildetotM})
on setting $\bfr_j = \bfr_{j0} + \bm{\xi}_j(0) + (\bfv_j + \dot{\bm{\xi}}_j(0)) t$
for all $j$'s in the latter and excluding the $l$-th term. We define the two mismatches to ballistic values
\begin{eqnarray}
  \Delta \bm{\xi}_l (t)
  & = &
  \bm{\xi}_l (t) - \bm{\xi}_l(0) - \dot{\bm{\xi}}_l(0) t  ,
  \label{Delta_xi}
  \\
  \Delta \tilde{\varphi}_l (\bfm,t)
  & = &
  \tilde{\varphi}_l (\bfm,t) - \phi_l^{(\rm{bal})} (\bfm,t) .
  \label{Delta_phi}
\end{eqnarray}

Until the end of section \ref{DSLD}, we consider cases where all the $\bm{\xi}_l$'s are small.
So we approximate $\Delta \tilde{\varphi}_l(\bfm)$ by its expansion
to first order in the $\Delta \bm{\xi}_l$'s (Approximation 1, discussed in \ref{DA1})
\begin{equation}
  \Delta \tilde{\phi}_l (\bfm,t)
  =
  \sum_{j \in S;j \neq l} \delta \tilde{\phi}_{j} (\bfm,t),
\label{phitildn}
\end{equation}
with
\begin{equation}
  \delta \tilde{\phi}_{j} (\bfm,t)
  =
  \frac{\rmi e}{\epsilon_0 k_{\bfm}^2} \exp [- \rmi \bfk_{\bfm} \cdot \bfr_{j}^{(0)}(t)] \ \bfk_{\bfm} \cdot \Delta \bm{\xi}_j  ,
\label{phitildnj}
\end{equation}
the contribution of particle $j$. This provides a first order approximation in the $\Delta \bm{\xi}_l$'s to $\tilde{\varphi}_l$, namely
\begin{equation}
  \tilde{\phi}_l (\bfm,t)
  =
  \phi_l^{(\rm{bal})} (\bfm,t) + \Delta \tilde{\phi}_l (\bfm,t) .
\label{phitildmt}
\end{equation}
We further consider $\varphi$ to be small, and the $\bm{\xi}_l$'s to be of the order of $\varphi$ (Approximation 2).
At lowest order, the particles dynamics defined by Eq.\ (\ref{rsectot}) is then given by
\begin{equation}
  \Delta \ddot{\bm{\xi}}_l
  =
  \ddot{\bm{\xi}}_l
  =
  \frac{\rmi e}{L^3 m_\rme} \sum_{\bfn} \bfk_{\bfn} \
    \tilde{\phi}_l(\bfn,t) \exp(\rmi \bfk_{\bfn} \cdot \bfr_l^{(0)}(t)).
\label{delrsec}
\end{equation}

We now use the time Laplace transform which maps a function $ g(t)$
to $\widehat{g}(\omega) = \int_0^{\infty}  g(t) \exp(\rmi \omega t) \rmd t$
(with $\omega$ complex). Since the arguments of functions are spelled explicitly,
from now on we omit diacritics for the Fourier-Laplace transformed quantities.
Note that $g(\bfm, \omega)^* = g(- \bfm, - \omega^*)$ for real $\bfm$ and complex $\omega$ if $g(\bfr, t)$ is real-valued.
The Laplace transform of Eq.\ (\ref{delrsec}) is
\begin{equation}
  \omega^2 \bm{\xi}_l(\omega)
  = - \frac{\rmi e}{L^3 m_\rme} \sum_{\bfn}
                \bfk_{\bfn} \exp(\rmi \bfk_{\bfn} \cdot \bfr_{l0})
                  \ \phi_l(\bfn,\omega + \omega_{\bfn,l})
                  + \rmi \omega \bm{\xi}_l(0) - \dot{\bm{\xi}}_l(0),
\label{rLapl}
\end{equation}
or
\begin{equation}
  \omega^2 \Delta \bm{\xi}_l(\omega)
  = - \frac{\rmi e}{L^3 m_\rme} \sum_{\bfn}
                \bfk_{\bfn} \exp(\rmi \bfk_{\bfn} \cdot \bfr_{l0})
                  \ \phi_l(\bfn,\omega + \omega_{\bfn,l}) ,
\label{rLaplDelta}
\end{equation}
where $\omega_{\bfn,l} = \bfk_{\bfn} \cdot \bfv_{l}$
comes from the time dependence of $\bfr_l^{(0)}$, as defined by Eq.\ (\ref{rl0}), in the exponent of Eq.\ (\ref{delrsec}).
Equation (\ref{rLapl}) and the Laplace transform of Eqs\ (\ref{phitildn}) and (\ref{phitildnj}) then yield
\begin{equation}
  % \fl  
 k_{\bfm}^2 \Delta \phi_l(\bfm,\omega)
  = \frac{\omega_{\rmpp}^2}{N}
      \sum_{\bfn} \bfk_{\bfm} \cdot \bfk_{\bfn}
    \sum_{j \in S;j \neq l} \frac{\phi_j(\bfn,\omega + \omega_{\bfn,j} - \omega_{\bfm,j})}
                                                 {(\omega - \omega_{\bfm,j})^2} \exp[\rmi (\bfk_{\bfn}-\bfk_{\bfm}) \cdot \bfr_{j0}] ,
\label{phihatnf}
\end{equation}
where $\omega_{\bfm,j}$ comes from the time dependence of $\bfr_j^{(0)}$ in the exponent of Eq.\ (\ref{phitildnj}),
and $\omega_{\rmpp} = [e^2 n / (m_\rme \epsilon_0)]^{1/2}$ is the plasma frequency.

Summing Eq.\ (\ref{phihatnf}) over $l = 1,... N$, using Eq.\ \ (\ref{phitildmt}), and dividing by $N-1$  yields
\begin{eqnarray}
% \fl  
&&
   k_{\bfm}^2\phi(\bfm,\omega)
 - \frac{\omega_{\rmpp}^2}{N}
 \sum_{\bfn} \bfk_{\bfm} \cdot \bfk_{\bfn}
  \ \sum_{j \in S} \frac{\phi(\bfn,\omega + \omega_{\bfn,j} - \omega_{\bfm,j})}{(\omega - \omega_{\bfm,j})^2}
                             \exp[\rmi (\bfk_{\bfn}-\bfk_{\bfm}) \cdot \bfr_{j0}]
  \nonumber \\
% \fl  
 & = &
  k_{\bfm}^2 \phi^{(\rm{bal})}(\bfm,\omega) ,
\label{phihat}
\end{eqnarray}
where $\phi(\bfm,\omega)$ and $\phi^{(\rm{bal})}(\bfm,\omega) $ are respectively
$\phi_l (\bfm,\omega)$ and $\phi_l^{(\rm{bal})}(\bfm,\omega) $ complemented with the missing $l$-th term.
More explicitly,
\begin{equation}
  \phi^{(\rm{bal})}(\bfm,\omega) =
  \sum_{j \in S} \delta \phi_j^{(\rm{bal})}(\bfm,\omega)
  ,
\label{phi0hat}
\end{equation}
where
\begin{equation}
% \fl  
  \delta \phi_j^{(\rm{bal})}(\bfm,\omega)
  = - \frac{\rmi e}{\epsilon_0 k_{\bfm}^2}
      \frac{\exp[- \rmi \bfk_{\bfm}  \cdot (\bfr_{j0} + \bm{\xi}_j(0))]}
             {\omega -\bfk_{\bfm}  \cdot (\bfv_j + \dot {\bm{\xi}}_j(0))}
  = - \frac{\rmi e}{\epsilon_0 k_{\bfm}^2}
      \frac{\exp[- \rmi \bfk_{\bfm}  \cdot \bfr_j(0)]}
             {\omega -\bfk_{\bfm}  \cdot \dot \bfr_j(0)} .
\label{phij0hat}
\end{equation}
The division by $N-1$ and not by $N$ is a direct consequence of the exclusion of self-interaction terms in the $\phi_l$'s.
Equation (\ref{phihat}) is the fundamental equation of this paper.
\emph{This equation is of the type ${\mathcal{E}} \phi=$ source term},
where ${\mathcal{E}}$ is a linear operator,
acting on the infinite dimensional array whose components are all the Doppler shifted $\phi(\bfm,\omega)$'s.

%VVVVVVVVVVVVVVVVVVVVVVVVVVVVVVVVVVVVVVV
\section{Debye shielding, Langmuir waves and Landau damping}
\label{DSLD}
%VVVVVVVVVVVVVVVVVVVVVVVVVVVVVVVVVVVVVVV

We rewrite Eq. (\ref{phihat}) by separating its diagonal terms $\bfn = \bfm$ and its non-diagonal terms $\bfn \neq \bfm$ as
\begin{eqnarray}
% \fl  
  && \epsilon_\rmd(\bfm,\omega) \, \phi(\bfm,\omega)
\nonumber\\
% \fl
  & = & \phi^{(\rm{bal})}(\bfm,\omega) + \frac{\omega_{\rmpp}^2}{N}
 \sum_{\bfn \neq \bfm} \frac{\bfk_{\bfm} \cdot \bfk_{\bfn}}{k_{\bfm}^2}
  \ \sum_{j \in S} \frac{\phi(\bfn,\omega + \omega_{\bfn,j} - \omega_{\bfm,j})}{(\omega - \omega_{\bfm,j})^2}
                             \exp[\rmi (\bfk_{\bfn}-\bfk_{\bfm}) \cdot \bfr_{j0}],
\label{phihatdiNdi}
\end{eqnarray}
where
\begin{equation}
  \epsilon_\rmd(\bfm,\omega)
  = 1 - \frac{\omega_{\rmpp}^2}{N}
  \sum_{j \in S} \frac{1}{(\omega - \omega_{\bfm,j})^2}.
\label{epsdiscr}
\end{equation}

%VVVVVVVVVVVVVVVVVVVVVVVVVVVVVVVVVVVVVVV
\subsection{Shielded Coulomb potential}
\label{SCP}
%VVVVVVVVVVVVVVVVVVVVVVVVVVVVVVVVVVVVVVV

We now approximate the granular distribution
$F_0 = \sum_{j \in S} \delta(\bfr - \bfr_{j0}) \  \delta(\bfv -  \bfv_{j})$
with a position and velocity distribution function $f_0(\bfr,\bfv)$
that is \emph{continuous} in $\bfr$, such that distribution $f_0-F_0$ yields a negligible contribution
when applied to space dependent functions which vary slowly on the scale of the inter-particle distance.
We further assume $f_0$ to be spatially uniform, and we write it $f_0(\bfv)$
(normalized with $\iint f_0(\bfv)\, \rmd^3 \bfr \, \rmd^3 \bfv = L^3 \int  f_0(\bfv)\, \rmd^3 \bfv = N$).

We replace the discrete sums over particles with integrals over $f_0$ (Approximation 3).
This has two consequences : operator $\mathcal{E}$ becomes diagonal
with respect to both $\bfm$ and $\omega$ (which is complex), and $\epsilon_\rmd(\bfm,\omega)$ becomes
\begin{equation}
  \epsilon(\bfm,\omega)
  = 1 - \frac{\omega_{\rmpp}^2 L^3}{N}
     \int \frac{f_0(\bfv) }{(\omega - \bfk_{\bfm}  \cdot \bfv)^2} \ \rmd^3 \bfv.
\label{eps}
\end{equation}
Therefore, Eq.\ (\ref{phihatdiNdi}) becomes
\begin{equation}
  \epsilon(\bfm,\omega) \, \Phi(\bfm,\omega)
  = \phi^{(\rm{bal})}(\bfm,\omega),
\label{phihatL}
\end{equation}
where $\Phi$ is the smoothed version of $\phi$  (approximating $\varphi$) resulting from
Approximations 1 to 3.

This shows that the smoothed self-consistent potential $\Phi$
is determined by the response function $\epsilon(\bfm,\omega)$,
viz.\  the classical plasma dielectric function.
One readily checks this for a cold plasma~:
then $\epsilon(\bfm,\omega) = 1 - {\omega_{\rmpp}^2}/{\omega^2}$,
where $\omega_{\rmpp}$ is now computed with the plasma density $n = N/L^3$.
The classical expression involving $\partial f_0 / \partial \bfv$
obtains by a mere integration by parts if $f_0$ is differentiable.

As a result of Eq.\ (\ref{phi0hat}),
the part of $\Phi(\bfm,\omega)$ generated by particle $j$
is $\delta \Phi_j(\bfm,\omega)
= \delta \phi_j^{(\rm{bal})}(\bfm,\omega)/\epsilon(\bfm,\omega)$.
By inverse Fourier-Laplace transform, after some transient discussed later,
the potential due to particle $j$ is the sum of two parts~:
one due to the pole $\omega = \bfk_{\bfm}  \cdot \dot \bfr_j(0)$,
and one to the poles of $1/\epsilon(\bfm,\omega)$.

The first part is the shielded Coulomb potential \cite{Gasio,Bal,Rost}
\begin{equation}
  \delta \Phi_j (\bfr, t)
  = \delta \Phi(\bfr - \bfr_j(0) - \dot{\bfr}_j(0) t,\dot{\bfr}_j(0)),
\label{phij}
\end{equation}
where
\begin{equation}
  \delta \Phi (\bfr,\bfv)
  = - \frac{e}{L^3 \epsilon_0} \sum_{{\bfm} \neq {\mathbf{0}}}
      \frac{\exp(\rmi \bfk_{\bfm} \cdot \bfr)}
           { k_{\bfm}^2 \, \epsilon(\bfm,\bfk_{\bfm} \cdot \bfv + \rmi \varepsilon)}
\label{phi}
\end{equation}
with the usual $\rmi \varepsilon$ prescription resulting from inverting the Laplace transform, as
the integral in Eq.\ (\ref{eps}) is undefined for the real-valued
$\omega = \bfk_{\bfm} \cdot \bfv$.
Therefore, after this transient, except for possible Langmuir waves,
\emph{the dominant contribution to the full potential in the plasma turns out to be
the sum of the shielded Coulomb potentials of individual particles}
located at their ballistic positions computed with their initial positions and velocities.
This property enables a calculation of the collisional transport for all impact parameters \cite{EED14}.

Let $\lambda_\rmD = [\epsilon_0 k_{\rm{B}} T / (n e^2)]^{1/2}
     = [k_{\rm{B}} T / m_\rme]^{1/2} \omega_{\rmpp}^{-1}$
be the Debye length, where $k_{\rm{B}}$ is the Boltzmann constant and $T$ the temperature.
The wavenumbers resolving scale $\| \bfr\|$ are such that $k_{\bfm} \|\bfr\| \gtrsim 1$.
Shielding involves scales on the order of $\lambda_\rmD$.
The transient is given by the zeros of $\epsilon(\bfm,\omega)$.
For shielding scales, these zeros correspond to a strong damping over time scales on the order of the plasma period.
Therefore, the transient is damped after such a period, as estimated in statement \ref{MR}.(\ref{claim-5}).

If $\|\bfr\| \ll \lambda_\rmD$, the corresponding wavenumbers are such
that $k_{\bfm} \lambda_\rmD \gg 1$.
Therefore, there is no shielding for $\|\bfr\| \ll \lambda_\rmD$,
since $\epsilon(\bfm,\bfk_{\bfm} \cdot \bfv) - 1
  \simeq - [v_{\mathrm{th}} / (\lambda_\rmD \, \bfk_{\bfm} \cdot \bfv)]^2
  \approx - (k_{\bfm} \lambda_\rmD)^{-2}$
where $v_{\mathrm{th}} = \lambda_\rmD \, \omega_{\rmpp}$ is the thermal velocity.

%VVVVVVVVVVVVVVVVVVVVVVVVVVVVVVVVVVVVVVV
\subsection{Langmuir waves and Landau damping}
\label{STSP}
%VVVVVVVVVVVVVVVVVVVVVVVVVVVVVVVVVVVVVVV

We now consider the part of the potential due to particle $j$ provided by the two dominant (Landau) poles of $1/\epsilon(\bfm,\omega)$.
These two poles correspond to one wave propagating in the direction of $\bfk_{\bfm}$ and one propagating in the opposite direction.
For simplicity, we retain only the first one (which we call $\omega_{\bfm}$), and associate the other one to $\bfk_{- \bfm}$.
By retaining this pole in the inverse Laplace transform of Eq.\ (\ref{phihatL})
and using Eqs (\ref{phiInv}) and (\ref{phi0hat})-(\ref{phij0hat}), one finds that particle $j$ brings a contribution to the
wave with frequency $\omega_{\bfm}$
\begin{equation}
  \Phi_{j \, \bfm}(\bfr, t)
  =
  - \frac{e}{\epsilon_0 k_{\bfm}^2 L^3 \, \epsilon'_\bfm}
      \frac{\exp[\rmi (\bfk_{\bfm} \cdot (\bfr - \bfr_j(0)) - \omega_{\bfm} t)]}
             {\omega_{\bfm} -\bfk_{\bfm}  \cdot \dot \bfr_j(0)} + \mathrm{c.\,c.} \, ,
\label{Phijt}
\end{equation}
where ``c.\,c.'' means complex conjugate, and $\epsilon'_\bfm =
  \frac{\partial \epsilon}{\partial \omega} (\bfm, \omega_{\bfm})$.

We now approximate the granular distribution
$F = \sum_{j \in S} \delta(\bfr - \bfr_j(0)) \  \delta(\bfv -  \dot \bfr_j(0))$
with a smooth position and velocity distribution function $f(\bfr,\bfv)$ in the sum over $j$ of the $\Phi_{j \, \bfm}(\bfr, t)$'s.
This sum yields the total amplitude of the wave with frequency $\omega_{\bfm}$
\begin{equation}
\Phi_{\bfm}(\bfr, t)
  = - e \ \frac{ \exp[\rmi (\bfk_{\bfm} \cdot \bfr - \omega_{\bfm} t)]}{\epsilon_0 k_{\bfm}^2 L^3 \, \epsilon'_\bfm}
    \int
    \frac{f(\bfm,\bfv)}
         {\omega_{\bfm} -\bfk_{\bfm} \cdot \bfv} \
     \rmd^3 \bfv
     + \mathrm{c.\,c.} \, ,
     \label{Phift}
\end{equation}
which is the Vlasovian result for this wave.

Since Eqs (\ref{eps}) and (\ref{Phift}) do not involve derivatives of the distribution functions,
they also enable computing Langmuir waves induced by an initial perturbation
in the case where they are non-differentiable (for instance a two-stream one).

The other poles of $1/\epsilon(\bfm,\omega)$ correspond to a strong damping,
and bring again a transient which vanishes over time scales on the order of the plasma period.
We notice that $1/\epsilon(\bfm,\omega)$ converges to 1 when $\omega$ goes to infinity.
Therefore, the inverse Laplace transform of $1/\epsilon(\bfm,\omega)$ has a singular part, the Dirac distribution $\delta (t)$.
As a result, the convolution involved in the inverse Laplace transform of Eq.\ (\ref{phihatL})
brings the ballistic potential $\phi^{(\rm{bal})}(\bfm,t)$ which dominates transiently, as expected.

%VVVVVVVVVVVVVVVVVVVVVVVVVVVVVVVVVVVVVVV
\subsection{Vlasovian formula for the ballistic potential}
\label{Vfor}
%VVVVVVVVVVVVVVVVVVVVVVVVVVVVVVVVVVVVVVV

We now recover the usual textbook expressions by introducing $f(\bfr,\bfv)$
before performing the inverse Laplace transform providing $\Phi(\bfm, t)$.
On neglecting the $\bm{\xi}_j(0)$'s and the $\dot {\bm{\xi}}_j(0)$'s to lowest
order in Eq.\ (\ref{phij0hat}), this yields
a  $\Phi^{(\rm{bal})}(\bfm, t)$
whose Laplace transform is
\begin{equation}
  \Phi^{(\rm{bal})}(\bfm,\omega)
  = - \frac{\rmi e}{\epsilon_0 k_{\bfm}^2} \int
    \frac{f(\bfm,\bfv)}
         {\omega -\bfk_{\bfm} \cdot \bfv} \
     \rmd^3 \bfv,
\label{phi0hatcg}
\end{equation}
the smoothed version of the actual shielded potential in the plasma.
This shows that this second continuous approximation makes Eq.\ (\ref{phihatL})
\emph{to become the expression including initial conditions in Landau contour calculations of Langmuir wave growth or damping},
usually obtained by linearizing Vlasov equation and using Fourier-Laplace transform, as described in many textbooks. In general, textbooks do not dwell upon $\Phi^{(\rm{bal})}(\bfm,\omega)$. The $N$-body description reveals that it is the continuous limit of a granular source term bringing not only the excitation of Langmuir waves, but also the Debye-shielded potential of the particles.

%VVVVVVVVVVVVVVVVVVVVVVVVVVVVVVVVVVVVVVV
\subsection{Comparison with kinetic approaches}
\label{Cwka}
%VVVVVVVVVVVVVVVVVVVVVVVVVVVVVVVVVVVVVVV

It is interesting to compare the above derivation with that presented by classical textbooks
when they start with the $N$-body description to derive both Debye shielding
and the combination of Eqs\ (\ref{phihatL}) and (\ref{phi0hatcg}).
Debye shielding is exhibited in the equilibrium pair correlation function computed
after deriving the first two equations of the BBGKY hierarchy (see  e.g.\ ch.\ 12 of \cite{BoydSan}).
The combination of Eqs\ (\ref{phihatL}) and (\ref{phi0hatcg}) is obtained independently
by linearizing Vlasov equation about a uniform velocity distribution function,
and by using the Fourier-Laplace transform.
A prerequisite is the derivation of Vlasov equation by either of two approaches~:
a mean-field derivation \cite{NeunzertWick,Neunzert84, Dobru,BraunHepp,Spohn,Kie14,ElsVla}
that does not involve statistical arguments,
or the BBGKY hierarchy that involves statistical arguments
and starts with the Liouville equation (see e.g.\  \cite{Nicholson}).
Rigorously speaking, these two derivations do not give the same definition to the velocity distribution function.

The present derivation Laplace-transforms in time the linearized dynamics of a \emph{single realization} of the $N$-body system.
This yields Eq.\ (\ref{phihat}) which keeps the full graininess of the system.
A first continuous approximation involving a velocity distribution function yields Eqs (\ref{phij})-(\ref{phi}),
and a second one yields Eq.\ (\ref{phi0hatcg}) combined with Eq.\ (\ref{phihatL}).
This provides a short connection between these equations and the underlying $N$-body problem.
In this derivation, the continuous velocity distribution is introduced after particle dynamics has been taken into account,
and not before, as occurs when kinetic equations are used.
This avoids addressing the issues of the exact definition of the continuous distribution for a given realization of the plasma,
and of the uncertainty as to the way the continuous dynamics departs from the actual $N$-body one \cite{NeunzertWick,Neunzert84,Dobru,BraunHepp,Spohn,Kie14,ElsVla}.
Introducing the continuous velocity distribution after particle dynamics has been taken into account may bring important new physics~:
it exhibited that the true (granular) plasma goes beyond the Vlasovian saturation for the growth of a single wave due to a warm beam \cite{Firpo}.

%VVVVVVVVVVVVVVVVVVVVVVVVVVVVVVVVVVVVVVV
\subsection{More rigorous derivation and range of validity}
\label{Mrdrv}
%VVVVVVVVVVVVVVVVVVVVVVVVVVVVVVVVVVVVVVV

Up to this point, linearization was applied without questioning its range of validity.
However, for any finite value of $\| \Delta \bm{\xi}_j \|$,
the linearization of Eq.\ (\ref{phitildnj}) is justified for finite values of $k_{\bfm}$ only.
Fortunately, we notice that the above introduction of Debye shielding and of Langmuir waves
involves scales larger than, or of the order of $\lambda_\rmD$.
As a result, we do not need a correct description of the dynamics due to the large wavenumbers,
i.e.\ due to the $1/r$ singularity of the Coulomb potential.
Therefore, we may restrict the expansion of Eq.\ (\ref{phiInv}) to $\bfm$'s such that $k_{\bfm} b_{\mathrm{smooth}} \leq 1$,
where $b_{\mathrm{smooth}} \ll \lambda_\rmD$.
Then the summation in Eq.\ (\ref{phihat}) is bounded similarly, and we may consider finite $\| \Delta \bm{\xi}_j \|$'s,
much smaller than $b_{\mathrm{smooth}}$ though.
We notice that a similar smoothing of the $1/r$ singularity of the Coulomb potential
(or some penalization of very close approaches \cite{Kie14})
is performed in the mean-field derivation of the Vlasov equation \cite{NeunzertWick,Neunzert84, Dobru,BraunHepp,Spohn,ElsVla}.

Using the smoothed version of Coulomb potential recovers the shielded Coulomb potential of Eqs (\ref{phij})-(\ref{phi})
down to a distance $\sim b_{\mathrm{smooth}}$ from the particle of interest.
Since $b_{\mathrm{smooth}} \ll \lambda_\rmD$, a part of the $1/r$ dependence of the genuine Coulomb potential is recovered,
and can be matched with the central $1/r$ dependence for closer distances.

\ref{DSmoo} considers a velocity distribution corresponding to a large number of monokinetic beams.
It shows that, if the inter-particle distance $d = n^{-1/3}$ verifies condition
\begin{equation}
d \ll b_{\mathrm{smooth}},
\label{dllbsmoo}
\end{equation}
in Eq.\ (\ref{phihatdiNdi}) the non-diagonal terms vanish,
which is the main simplification brought by the summation over the continuous distribution function at the beginning of section \ref{SCP}.
Since $b_{\mathrm{smooth}} \ll \lambda_\rmD$, then $d \ll \lambda_\rmD$ too,
which is equivalent to $\Lambda = n \lambda_\rmD^3 \gg 1$.
This shows that the presence of many particles in the Debye sphere justifies neglecting the non-diagonal terms in Eq.\ (\ref{phihatdiNdi}).

Then, this equation describes the response of a large number of monokinetic beams to the initial perturbation defined by $\phi^{(\rm{bal})}(\bfm,\omega)$.
This problem was considered by Dawson in 1960 for a one-dimensional plasma \cite{Dawson60}.
He showed that  $\epsilon_\rmd(\bfm,\omega)$ brings two beam modes per beam.
Their eigenfrequencies are pairs of complex conjugate values for $\omega$,
whose imaginary parts vanish when the spacing of the beam velocity decreases~:
this makes these modes analogous to the van Kampen modes.
Landau damping is recovered by phase mixing of these modes.
Dawson's analysis can be directly used by restricting the plasma of interest to a one-dimensional OCP. Therefore, in this case
the use of the continuous distribution function at the beginning of section \ref{SCP} is justified when the number of particles in a Debye sphere becomes large.
The generalization of Dawson's calculation to the three-dimensional case a priori does not display any new conceptual difficulty, but is more involved and will not be given here.

The truncated Coulomb potential cannot correctly describe the encounters between particles with impact parameters smaller than $b_{\mathrm{smooth}}$,
which makes the description of the dynamics relevant for times shorter than the collision time
$\tau_{\rm{coll}} = 3(2 \pi)^{3/2} \Lambda /(\omega_\mathrm{p} \ln \Lambda) $,
as happens for the Vlasovian description.
Both descriptions are relevant for Langmuir waves,
since conditions $\Lambda \gg 1 $ and $\omega_{\rmpp} \tau_{\rm{coll}} \gg 1$ are equivalent.

%VVVVVVVVVVVVVVVVVVVVVVVVVVVVVVVVVVVVVVV
\section{Mediated interactions imply Debye shielding}
\label{MIIDS}
%VVVVVVVVVVVVVVVVVVVVVVVVVVVVVVVVVVVVVVV

In the above derivation of Debye shielding, using the Laplace transform of the particle positions does not provide an intuitive picture of this effect. We now show that such a picture is obtained directly from the mechanical description of microscopic dynamics with the full OCP Coulomb potential of Eq.\ (\ref{phitildetotM}) in real time. To compute the dynamics, we use Picard iteration technique.
From Eq.\ (\ref{rsectot}), $\bfr_l^{(n)}$, the $n$-th iterate for $\bfr_l$, is computed from
\begin{equation}
  \ddot{\bfr}^{(n)}_l
  = \frac{e}{m_\rme} \nabla \varphi_l^{(n-1)}(\bfr^{(n-1)}_l),
\label{rsecn}
\end{equation}
where $\varphi_l^{(n-1)}$ is computed by the inverse Fourier transform of Eq.\ (\ref{phitildetotM}) with the $\bfr_j$'s substituted with the $\bfr_j^{(n-1)}$'s. The iteration starts with the ballistic approximation of the dynamics defined by Eq.\ (\ref{rl0}), and the actual orbit of Eq.\ (\ref{rsectot}) corresponds to $n \rightarrow \infty$. Let $\bm{\xi}_l^{(n)} = \bfr_l^{(n)} - \bfr_l^{(0)}$ be the mismatch of the position of particle $l$ with respect to the ballistic one at the $n$-th iterate, viz.\ the effect of Coulomb interactions to that order of iterations~; we assume all $\bm{\xi}_l^{(0)}$'s and $\dot{\bm{\xi}}^{(0)}_l$'s to vanish identically. It is convenient to write Eq.\ (\ref{rsecn}) as
\begin{equation}
 \ddot{\bm{\xi}}^{(n)}_l
  = \sum_{j \in S;j \neq l} \ddot{\bm{\xi}}^{(n)}_{lj},
\label{rsecnAccl}
\end{equation}
with
\begin{equation}
  \ddot{\bm{\xi}}^{(n)}_{lj}
  = \bfa_{\rm{C}}(\bfr_l^{(n-1)}-\bfr_j^{(n-1)})
\label{rsecnAcclj}
\end{equation}
 and
\begin{equation}
  \bfa_{\rm{C}}(\bfr)
  =
  \frac{\rmi e^2}{\epsilon_0 m_\rme L^3}
  \sum_{{\bfm} \neq {\mathbf{0}}}
      k_{\bfm}^{-2} \, \bfk_{\bfm}
      \exp(\rmi \bfk_{\bfm} \cdot \bfr).
\label{phiCbPer}
\end{equation}
Let $\bm{\xi}^{(n)}_{lj}
    = \int_0^t \int_0^{t'} \ddot {\bm{\xi}}^{(n)}_{lj} (t'') \, \rmd t'' \rmd t' = \int_0^t (t-t'') \, \ddot {\bm{\xi}}^{(n)}_{lj} (t'') \, \rmd t''$.
For $n \geq 2$, one finds
\begin{equation}
  \ddot{\bm{\xi}}^{(n)}_l
  = \sum_{j \in S;j \neq l}
       [\ddot{\bm{\xi}}^{(1)}_{lj} + M_{lj}^{(n-1)}
        + 2 \nabla \bfa_{\rm{C}}(\bfr_l^{(0)}-\bfr_j^{(0)}) \cdot \bm{\xi}_{lj}^{(n-1)}] + O(a^3),
\label{rsecnAccDev2}
\end{equation}
where $a$ is the order of magnitude of the total Coulombian acceleration, and
\begin{eqnarray}
M_{lj}^{(n-1)}
  & = & \nabla \bfa_{\rm{C}}(\bfr_l^{(0)}-\bfr_j^{(0)})
      \cdot [\bm{\xi}_{l}^{(n-1)} - \bm{\xi}_{j}^{(n-1)} - 2 \bm{\xi}_{lj}^{(n-1)}]
  \\
  & = & \nabla \bfa_{\rm{C}}(\bfr_l^{(0)}-\bfr_j^{(0)})
      \cdot \sum_{i \in S; i \neq l,j} (\bm{\xi}_{li}^{(n-1)} + \bm{\xi}_{ij}^{(n-1)})   ,
\label{Mlj}
\end{eqnarray}
where the second expression takes into account that $\bfa_{\rm{C}}(\bfr)$ is anti-symmetrical in $\bfr$.
The latter expression displays $\bm{\xi}_{ij}^{(n-1)}$ which is the deflection of particle $i$ by particle $j$.
It shows the bare Coulomb acceleration of particle $l$ due to particle $j$
is modified by the following process~:
particle $j$ modifies the motion of all other particles,
so that the action of the latter ones on particle $l$ is modified by particle $j$.
Therefore $M_{lj}^{(n-1)} $ \emph{is the acceleration of particle $l$ due to particle $j$ mediated by all other particles}.
The last term in the bracket in Eq.\ (\ref{rsecnAccDev2}) accounts for the fact
that both particles $j$ and $l$ are shifted with respect to their ballistic positions.
Both $M_{lj}^{(n-1)} $ and this last term are anti-symmetrical with respect to the labels $j$ and $l$
as $\nabla \bfa_{\rm{C}}(\bfr)$ is an even function of $\bfr$.

Since the shielded potential of the previous paragraph was found by first order perturbation theory,
it is felt in the acceleration of particles computed to second order.
This acceleration is provided by Eq.\ (\ref{rsecnAccDev2}) for $n=2$.
Therefore its term in brackets is the shielded acceleration of particle $l$ due to particle $j$.
As a result, though the summation runs over all particles, its effective part is only due
to particles $j$ typically inside the Debye sphere (with radius $\lambda_\rmD$) about particle $l$.
Starting from the third iterate of the Picard scheme, the effective part of the summation in Eq.\ (\ref{rsecnAccDev2})
ranges inside this Debye sphere, since the $\bm{\xi}_{lj}^{(n-1)}$'s are then computed with a shielded acceleration.
This approach clarifies the mechanical background of the calculation of shielding using the equilibrium pair correlation function
which shows shielding to result from the correlation of two particles occurring through the action of all the other ones
(see e.g.\ section 12.3 of \cite{BoydSan}).
In Eq.\ (\ref{Mlj}), the compound effect of the $\bm{\xi}_{ij}^{(n-1)}$'s,
the deflection of particle $i$ by particle $j$, is to diminish the negative charge inside a sphere centered on particle $j$.
This yields the interpretation of shielding given in statement \ref{MR}.(\ref{claim-5}) using Gauss' theorem.

%TTTTTTTTTTTTTTTTTTTTTTTTTTTTTTTTTTTTTT
\section{Wave-particle dynamics}
\label{WPDRT}
%TTTTTTTTTTTTTTTTTTTTTTTTTTTTTTTTTTTTTT

Section \ref{STSP} enables the calculation of Langmuir waves excited by a given initial perturbation
after a continuous approximation of the true granular distribution.
To describe Langmuir waves with discrete particles, we still consider that the $\bfr_{l0}$'s are uniformly distributed,
and we allow for non-zero $ \bm{\xi}_j(0)$'s and $\dot{\bm{\xi}}_j(0)$'s for the $\bm{\xi}_j$'s of section \ref{FEP}.
Therefore, in the formulas of section \ref{DSLD}, the $\bfr_j(0)$'s
and $\dot{\bfr}_j(0)$'s are slightly shifted with respect to the $\bfr_{j0}$'s
and $\bfv_j$'s, due to Langmuir waves.

Up to this point, we described Langmuir waves by a fully linear theory.
We now generalize the analysis of section \ref{FEP} to afford the description of nonlinear effects in wave-particle dynamics.
Indeed, resonant particles may experience trapping or chaotic dynamics,
which imply $\bfk_{\bfm} \cdot \bm{\xi}_l$'s of the order of $2 \pi$ or larger for wave $\bfk_{\bfm}$'s. For such particles,
it is not appropriate to make the linearizations leading to Eqs (\ref{phitildnj}) and (\ref{delrsec}).
However, these linearizations may still be justified for non-resonant particles over times
where trapping and chaos show up for resonant ones.
In order to keep the capability to describe the latter effects, we now split the set of $N$ particles into bulk and tail,
in the spirit of Refs \cite{OWM,OLMSS,AEE,EZE,EEbook}.
The bulk is defined as the set of particles which are not resonant with Langmuir waves.
We then perform the analysis of section \ref{FEP} for these $N_{\mathrm{bulk}}$ particles,
while keeping the exact contribution of the remaining $N_{\mathrm{tail}}$ particles to the electrostatic potential.
To this end, we number the tail particles from 1 to $N_{\mathrm{tail}}$,
the bulk ones from $N_{\mathrm{tail}}+1$ to $N = N_{\mathrm{bulk}} + N_{\mathrm{tail}}$,
and we call these respective sets of integers $S_{\mathrm{tail}}$ and $S_{\mathrm{bulk}}$. For $l \in S_{\mathrm{bulk}}$,
we now substitute Eq.~(\ref{phitildmt}) with
\begin{equation}
\phi_l (\bfm,t)
  = \phi_{\mathrm{bulk}}^{(\rm{bal})}(\bfm,t) +
    \sum_{j \in S_{\mathrm{bulk}};j \neq l} \ \delta \phi_{j} (\bfm,t)
    + \frac{N_{\mathrm{bulk}}  - 1}{N_{\mathrm{bulk}}} U(\bfm,t) ,
\label{phitildnT}
\end{equation}
where
\begin{equation}
U(\bfm, t)
= -\frac{e N_{\mathrm{bulk}}}{\epsilon_0 k_{\bfm}^2 (N_{\mathrm{bulk}}-1)}
   \sum_{j \in S_{\mathrm{tail}}} \exp(- \rmi \bfk_{\bfm} \cdot \bfr_j),
\label{U}
\end{equation}
and from now on, all quantities with subscript ``$\mathrm{bulk}$" are those of previous
sections computed with the bulk particles only.
In the r.h.s.\ of Eq.\ (\ref{phitildnT}), the third term vanishes if $N_{\mathrm{tail}} = 0$. We now perform the calculation of section \ref{FEP} on substituting the previous summations with index running from 1 to $N$ by ones where the index runs over $S_{\mathrm{bulk}}$, while keeping the exclusion of $j=l$ where indicated.
The previous division by $N-1$ preceding Eq.~(\ref{phihat}) is now a division by $ N_{\mathrm{bulk}} - 1$. This yields
\begin{eqnarray}
% \fl  
  &&
   k_{\bfm}^2\phi(\bfm,\omega)
  - \frac{\omega_{\rmpp}^2}{N_{\mathrm{bulk}}}
 \sum_{\bfn} \bfk_{\bfm} \cdot \bfk_{\bfn}
\ \sum_{j \in S_{\mathrm{bulk}}} \frac{\phi(\bfn,\omega + \omega_{\bfn,j} - \omega_{\bfm,j})}{(\omega - \omega_{\bfm,j})^2}
          \exp[\rmi (\bfk_{\bfn}-\bfk_{\bfm}) \cdot \bfr_{j0}]
\nonumber\\
% \fl  
  & = &  k_{\bfm}^2 \phi_{\mathrm{bulk}}^{(\rm{bal})}(\bfm,\omega)
           + k_{\bfm}^2 U(\bfm,\omega) .
\label{phihatU}
\end{eqnarray}
On performing for the bulk particles the same approximations as in sections \ref{SCP}
and \ref{STSP}, Eq.~(\ref{phihatL}) becomes
\begin{equation}
  \epsilon_{\mathrm{bulk}} (\bfm,\omega) \Phi(\bfm,\omega)
  = \phi_{\mathrm{bulk}}^{(\rm{bal})}(\bfm,\omega) + U(\bfm,\omega).
\label{phihatLU}
\end{equation}

%TTTTTTTTTTTTTTTTTTTTTTTTTTTTTTTTTTTTTT
\subsection{Amplitude equations}
\label{AmplitudeEqn}
%TTTTTTTTTTTTTTTTTTTTTTTTTTTTTTTTTTTTTT

For the scales much larger than $\lambda_\rmD$, the electric potential for the bulk is a superposition of Langmuir waves. The presence of tail particles slightly modifies these waves. We now derive an amplitude equation for the potential $\Phi(\bfm, t)$ of the wave with wavenumber $\bfm$ in a way similar to Refs \cite{OWM,OLMSS}. On the l.h.s.\ of Eq.\ (\ref{phihatLU}), the dielectric function defines
a Bohm-Gross type dispersion relation associated with plasma oscillations of the bulk,
and on the r.h.s.\ the tail particles provide $U$ with contributions oscillating in the Langmuir wave spectrum.
Therefore, the frequency of interest in Eq.\ (\ref{phihatLU}) for wavevector $\bfk_\bfm$ is close to the eigenfrequency $\omega_{\bfm}$ solving $\epsilon_{\mathrm{bulk}} (\bfm,\omega_{\bfm}) = 0$ corresponding to the wave propagating in the direction of $\bfk_\bfm$~; this frequency is real, since it is not resonant with the support of the bulk distribution function ($f (\bfv) = 0$ for all $\bfv$'s such that $\bfk_\bfm  \cdot \bfv = \omega_{\bfm}$).

Let $\Phi_{\mathrm{bulk}}(\bfm, \omega)$ be the solution
of Eq.~(\ref{phihatLU}) computed for $U(\bfm,\omega) = 0$, so that
\begin{equation}
\epsilon_{\mathrm{bulk}} (\bfm,\omega) (\Phi (\bfm,\omega) - \Phi_{\mathrm{bulk}} (\bfm,\omega))
  = U(\bfm,\omega) .
  \label{Fi-Fi}
\end{equation}
We are now looking for a $\Phi (\bfm,t)$ which is close to the potential of the Langmuir wave $\Phi_{\mathrm{bulk}}(\bfm,t) = A \exp(-\rmi \omega_{\bfm} t)$,
corresponding to $N_{\mathrm{tail}} = 0$, where $A$ is a constant equal to the sum for $j \in S_{\mathrm{bulk}}$ of
\begin{equation}
  A_j
  =
  - \frac{e}{\epsilon_0 k_{\bfm}^2 \epsilon'_\bfm} \
      \frac{\exp[- \rmi \bfk_{\bfm}  \cdot \bfr_j(0)]}
             {\omega_{\bfm} -\bfk_{\bfm}  \cdot \dot \bfr_j(0)} .
\label{Psijt}
\end{equation}
$\Phi_{j \, \bfm}(\bfr, t)$ in Eq. (\ref{Phijt}) is $A_j \exp[\rmi (\bfk_{\bfm} \cdot \bfr  - \omega_{\bfm} t)] + \mathrm{c.\,c.}$.
Let $g(\bfm,t) = \Phi( \bfm,t) / \Phi_{\mathrm{bulk}}(\bfm,t)$ with $g(\bfm,0) = 1$.
Therefore $\Phi(\bfm,\omega) = A\, g(\bfm, \omega - \omega_{\bfm})$,
which together with Eq.~(\ref{Fi-Fi}) yields
\begin{equation}
  A \, \epsilon_{\mathrm{bulk}} (\bfm,\omega_{\bfm} + \omega')  \,
      [g(\bfm,\omega') - \frac{\rmi}{\omega'}]
  = U(\bfm,\omega_{\bfm} + \omega'),
\label{eqf}
\end{equation}
where $\omega' = \omega - \omega_{\bfm}$.
If $N_{\mathrm{tail}} \ll N_{\mathrm{bulk}}$,
$g(\bfm,t)$ is a slowly evolving complex amplitude,
and the dominant part of $g(\bfm,\omega)$ is concentrated near zero.
This justifies Taylor-expanding $\epsilon_{\mathrm{bulk}}(\bfm,\omega_{\bfm} + \omega')$
about $\omega' = 0$ in Eq.~(\ref{eqf}),
which yields
$\frac{\partial \epsilon_{\mathrm{bulk}}(\bfm,\omega_{\bfm})}{\partial \omega} \omega'$
to lowest order.
Setting this into Eq.~(\ref{eqf}) and performing the inverse Laplace transform finally yields an amplitude equation for $\Phi(\bfm, t)$
\begin{equation}
% \fl
\frac{\partial \Phi(\bfm, t)}{\partial t}  + \rmi \omega_{\bfm} \Phi(\bfm, t)
  =
  \frac{\rmi e N_{\mathrm{bulk}}}
         {\epsilon_0 k_{\bfm}^2 (N_{\mathrm{bulk}}  - 1)
                      \frac{\partial \epsilon_{\mathrm{bulk}}}{\partial \omega}(\bfm,\omega_{\bfm})
         }
  \sum_{j \in S_{\mathrm{tail}}} \exp(- \rmi \bfk_{\bfm} \cdot \bfr_j)
\label{eqampl}
\end{equation}
similar to those in Refs \cite{OWM,OLMSS}.
The self-consistent dynamics of $M$ Langmuir waves and of the tail particles is ruled by Eq.\ (\ref{eqampl}) written for each wave
and by the equation of motion of these particles due to the $M$ waves,
\begin{equation}
  \ddot{\bfr}_j
  = \frac{\rmi e}{L^3 m_\rme} \sum
         \bfk_{\bfn} \Phi(\bfn,t)
            \exp(\rmi \bfk_{\bfn} \cdot \bfr_j) ,
\label{delrsecwv}
\end{equation}
where the summation runs over the indices $\bfn$ of the $M$ waves,
and the tail-tail interactions were neglected owing to the low density of the tail particles.

These two sets of equations generalize to three dimensions the self-consistent dynamics defined in Refs \cite{MK,AEE,EEbook}.
The study of this dynamics enables recovering Vlasovian linear theory with a mechanical understanding
(see \cite{E013,SenFest} for a synthetic presentation).
In particular, the reason why Landau damping cannot be a damped eigenmode is shown to be rooted deeply in Hamiltonian mechanics~:
a damped eigenmode must exist along with an unstable one, which is going to dominate with probability 1.
Landau damping is recovered as an analogue of van Kampen phase-mixing effect.
This phase-mixing in turn plays an essential role in the calculation of Landau instability in order to cancel the damped eigenmode
(section 3.8.3 of Ref.\  \cite{EEbook}). The self-consistent dynamics comes with an important bonus~:
it provides the information on particle dynamics in parallel with the wave's.
In particular, it reveals that both Landau damping and instability result
from the same synchronization mechanism of particles with waves.
This explains why there is a single formula for the rates of growth and damping,
though growth involves an eigenmode and damping a phase-mixing instead \cite{EZE,EEbook,EFields,Houches}.
This synchronization mechanism was indeed evidenced experimentally \cite{DovEsMa}.
In contrast, as yet the Vlasovian derivations of Landau damping do not provide the description of the corresponding individual evolution of particles,
which forces textbooks to come up with complementary mechanical models.
We point out that in Refs \cite{AEE,EEbook} the equivalent of Eqs~(\ref{eqampl})-(\ref{delrsecwv}) was obtained
without using any continuous approximation,
but by a direct mechanical reduction of degrees of freedom starting with the $N$-body problem.

%TTTTTTTTTTTTTTTTTTTTTTTTTTTTTTTTTTTTTT
\subsection{Spontaneous emission}
\label{SpontEmission}
%TTTTTTTTTTTTTTTTTTTTTTTTTTTTTTTTTTTTTT

For the sake of brevity, we do not develop here the full generalization of the analysis in Refs \cite{AEE,EEbook}~;
it is lengthy, but straightforward. However, since this analysis unifies spontaneous emission with Landau growth and damping,
we recall the result ruling the evolution of the amplitude of a Langmuir wave provided by perturbation calculations
where the right hand sides of Eqs~(\ref{eqampl})-(\ref{delrsecwv}) are considered small (of first order).
This is natural for Eq.~(\ref{eqampl}) since $N_{\mathrm{tail}} \ll N_{\mathrm{bulk}}$,
and for Eq.~(\ref{delrsecwv}) if the Langmuir waves have a low amplitude.
Let $J(\bfm,t) = \langle \Phi(\bfm,t) \Phi(- \bfm,t) \rangle$,
where the average is over the random initial positions of the tail particles (their distribution being spatially uniform).
Then a calculation to second order in $\Phi$ yields \begin{equation}
    \frac {\rmd J(\bfm,t)} {\rmd t}
    = 2 \gamma_{\bfm\,{\rm L}} J(\bfm,t) + S_{\bfm \, \mathrm{spont} }  ,
   \label{evampfinal}
\end{equation}
where $\gamma_{\bfm\,{\rm L}}$ is the Landau growth or damping rate given by
\begin{equation}
  \gamma_{\bfm \,{\rm L}} = \alpha_{\bfm}
  {\frac {\rmd f_{\rm{red}}} {\rmd v}}\left(\frac{\omega_{\bfm}}{ k_{\bfm}} ; \bfm\right)
  \label{SZ122}
\end{equation}
with
\begin{equation}
  \alpha_{\bfm}
  =
  \frac{\pi e^2 }{m_\rme \epsilon_0 \, k_{\bfm}^2  \frac{\partial \epsilon}{\partial \omega}(\bfm,\omega_{\bfm})}  ,
 \label{alj}
\end{equation}
and $f_{\rm{red}}$ is the reduced smoothed distribution function $f_{\rm{red}}(v; \bfm) = \iint f_0(v \hat{\bfk}_{\bfm} + \bfv_{\bot}) \ \rmd^2 \bfv_{\bot}$ where $\hat{\bfk}_{\bfm}$ is the unit vector along $\bfk_{\bfm}$ and $\bfv_{\bot}$ is the component of the velocity perpendicular to $\bfk_{\bfm}$~; $S_{\bfm \, \mathrm {spont} }$ is given by
\begin{equation}
  S_{\bfm \, \mathrm{spont} }
  =
  \frac{2 \alpha_{\bfm}^2}{\pi e^2  k_{\bfm} n}
     f_{\rm{red}}\left(\frac{\omega_{\bfm}}{ k_{\bfm}} ; \bfm\right)  ,
  \label{Spont}
\end{equation}
where $ n=N/L^3$ is the plasma density. $S_{\bfm \, \mathrm {\rm{spont}} }$ corresponds to the spontaneous emission of waves by particles and induces an exponential relaxation of the waves to the thermal level in the case of Landau damping (the analogue of what was found in \cite{EZE,EEbook}).
The second order calculation for the particles yields the diffusion and friction coefficients of the Fokker-Planck equation ruling the tail dynamics.
This equation corresponds to the classical quasilinear result, plus a dynamical friction term mirroring the spontaneous emission of waves by particles, as found in the one-dimensional case in Refs \cite{EZE,EEbook}.

An important aspect of the self-consistent dynamics defined
by Eqs~(\ref{eqampl})-(\ref{delrsecwv}) is that it enables to use the modern tools of nonlinear dynamics and chaos available for finite dimensional systems.
Let us consider two examples.
First, the van Kampen phase-mixing effect leading to Landau damping is nowadays a classical result of Vlasovian theory.
However, one may wonder whether nonlinear effects do not destroy these linear modes and the corresponding phase mixing.
Proving the innocuity of nonlinear effects is the equivalent
of deriving a Kolmogorov-Arnold-Moser (KAM) theorem for a continuous system (the Vlasov-Poisson one),
a tour de force which partly earned C.\ Villani the 2010 Fields medal \cite{MV,Villani}.
The same result for the above finite-dimensional self-consistent dynamics requires the standard KAM theorem only~:
it is much simpler to keep the genuine granularity of the plasma.

Second, consider a tail distribution function which is a plateau in both velocity and space
(this occurs for instance at the saturation of the bump-on-tail instability
in a particle description of the plasma).
Then the source term in Eq.~(\ref{eqampl}) vanishes, as well as mode coupling,
and the waves keep a fixed amplitude~:
the self-consistency of Eqs~(\ref{eqampl})-(\ref{delrsecwv}) is quenched,
even when particle dynamics is strongly chaotic in the plateau domain (see sec.\ 2.2 of \cite{BEEB}).
Then, it is possible to use the tools of 1.5 degree-of-freedom Hamiltonian chaos to compute the diffusion of particle velocities.
In particular, if chaos is strong enough, one may use a quasilinear diffusion coefficient (see section 2.2 of \cite{EE2}).
In a Vlasovian description, the bump-on-tail instability saturates with the previous plateau substituted with a very jagged distribution in both space and velocity, resulting from the chaotic stretching and bending of the initial beam-plasma distribution (the initial distribution $f_0$ is conserved along particle motion)~;
a plateau in velocity exists for the spaced-averaged distribution function only,
and a plateau in space exists for the velocity-averaged distribution only.

Third, when the initial amplitude of a Landau damped Langmuir wave is increased, there is a threshold above which the wave amplitude enters an oscillatory regime \cite{O'N}. The description of this phenomenon by the above self-consistent dynamics enables showing that this corresponds to a second order phase transition \cite{FirEl2}.

%TTTTTTTTTTTTTTTTTTTTTTTTTTTTTTTTTTTTTT
\section{Conclusion}
\label{Concl}
%TTTTTTTTTTTTTTTTTTTTTTTTTTTTTTTTTTTTTT

This paper provides a direct path from microscopic mechanics to Debye shielding, Landau damping, and wave-particle interaction.
This is performed by using Newton's second law for the $N$-body description, and standard tools of calculus.
In particular, sections \ref{FEP} to \ref{Vfor} provide the explicit, yet very compact derivation of formulas requiring about twenty pages
(e.g.\ ch.\ 4, secs 6.3--5 and sec.\ 9.2 in Ref.\ \cite{Nicholson} and sec.\ 12.3 in Ref.\ \cite{BoydSan})
in classical textbooks proceeding also explicitly from the $N$-body description.
This occurs thanks to a considerable simplification of the mathematical framework,
in particular because no probabilistic argument and no partial differential equation are used.

This paper also solves the following issue~:
How can each particle be shielded by all other ones, while all the plasma particles are in uninterrupted motion~?
This dynamical shielding turns out to be a mere consequence of the almost independent deflections of particles due to the Coulomb force.
Therefore, shielding and collisional transport are two related aspects of the repulsive deflections of electrons.

The new approach also shows that each particle brings two simultaneous contributions to the plasma potential~:
a short-range one, its shielded Coulomb potential, and a long-range one, the excitation of Langmuir waves.
However, these contributions need a time on the order of the plasma period to set in.
Therefore, the plasma behaves like a dielectric only after this time during which Coulomb deflections bring their cooperative contribution to this self-organization.
Paradoxically, Landau damping turns out to be correct because of what is usually called ``collisions".

The $N$-body dynamics has always been the ultimate reference in plasma textbooks.
Here it becomes a practical tool, which contributes to conform more strictly to the rules of inference from first principles~:
classical mechanics describes and unifies non-trivial aspects of the macroscopic dynamics of a many-body system,
and brings more insight in plasma self-organization.

One might contemplate applying the above mechanical approach to plasmas with more species,
or with a magnetic field, or where particles experience trapping and chaotic dynamics.
The first generalization sounds rather trivial, and the third one is under way,
at least in one dimension (see a pedestrian introduction in \cite{Houches} and more specific results in \cite{BEEB,BEEBEPS}).

J. Callen, Ph.~Choquard, L.~Cou\"edel, M.-C.~Firpo, W.~Horton, P.K.~Kaw, J.T.~Mendon\c{c}a, F.~Pegoraro, Y.~Peysson,
A.~Piel, H.~Schamel, D.~Zarzoso, J.-Z.~Zhu and an anonymous referee are thanked for very useful comments and references.
The Program Committee of the 41st EPS Conference on Plasma Physics is thanked
for providing one of us (DFE) with an extensive discussion of some aspects of the present paper.

%VVVVVVVVVVVVVVVVVVVVVVVVVVVVVVVVVVVVVVV
\appendix
%VVVVVVVVVVVVVVVVVVVVVVVVVVVVVVVVVVVVVVV
\section{Physical meaning of the linearization of the Coulomb potential (Approximation 1)}
\label{DA1}
%VVVVVVVVVVVVVVVVVVVVVVVVVVVVVVVVVVVVVVV

Approximation 1 in section \ref{FEP} corresponds to substituting the Coulomb potential in Eq.~(\ref{rsectot})
with $\phi_l = \phi_l^{(\rm{bal})} + \Delta \phi_l $ as defined by Eq.~(\ref{phitildmt}).
Using the definition of the ballistic approximation $\phi_l^{(\rm{bal})} (\bfm,t) $ and Eqs (\ref{phitildn})-(\ref{phitildnj}),
\begin{eqnarray}
%  \fl
  \phi_l (\bfm,t)
  &=&
  - \frac{e}{\epsilon_0 k_{\bfm}^2}
  \sum_{j \in S;j \neq l}
  \exp[- \rmi \bfk_{\bfm} \cdot \bfr_{j}^{(0)}(t)]
      \left(\rme^{- \rmi \bfk_{\bfm} \cdot (\bm{\xi}_j(0) + \dot{\bm{\xi}}_j(0) t)}
             - \rmi \bfk_{\bfm} \cdot \Delta \bm{\xi}_j \right)  .
\label{phitildl}
\end{eqnarray}
Taking into account that the $\bm{\xi}_j$'s are small, we may write
\begin{equation}
   \phi_l (\bfm,t)
   \simeq
   \sum_{j \in S;j \neq l} \delta \psi_j(\bfm,t),
\label{phitildlPsi}
\end{equation}
with
\begin{equation}
\delta \psi_j(\bfm,t) =
 - \frac{e}{\epsilon_0 k_{\bfm}^2} \exp(- \rmi \bfk_{\bfm} \cdot \bfr_{j}^{(0)}(t))(1 - \rmi \bfk_{\bfm} \cdot \bm{\xi}_j(t))  .
\label{phitildl2}
\end{equation}
The inverse Fourier transform of $\delta \psi_j(\bfm,t)$ verifies
\begin{equation}
  \lim_{L \to \infty} \delta \psi_{j} (\bfr,t)
  = - \frac{e}{4 \pi \epsilon_0 \| \bfr - \bfr_j^{(0)}\|}
- \frac{e \, \bm{\xi}_j \cdot (\bfr - \bfr_j^{(0)})}
    {4 \pi \epsilon_0 \| \bfr - \bfr_j^{(0)}\|^3}.
\label{Deltaphi}
\end{equation}
The $j$-th contribution to the approximate electric field turns out to be due to a particle located at $\bfr_{j}^{(0)}$ instead of $\bfr_j$,
and is made up of a Coulombian part and of a dipolar part with dipole moment $- e \, \bm{\xi}_j$.
The cross-over between these two parts occurs for $\| \bfr - \bfr_j^{(0)}\|$ on the order of $\| \bm{\xi}_j \|$,
i.e.\ when the distance to the ballistic particle $j$ is about the distance between the latter and the true particle $j$.
For larger values of $\| \bfr_l - \bfr_j^{(0)}\|$, the dipolar component is subdominant.
For smaller ones, it is dominant, but with a direction which is a priori uncorrelated with the Coulombian one ($(\bfr_l - \bfr_j^{(0)})$ is almost independent from $\bm{\xi}_j$).
Since the $\| \bm{\xi}_j \|$'s are assumed small, for particles having an almost uniform spatial distribution,
this dipolar part should be exceptionally dominant, as it corresponds to a very close encounter between particles.
As a result, the approximate electric field stays dominantly of Coulombian nature,
but with a small mismatch of the charge positions with respect to the actual ones.

%VVVVVVVVVVVVVVVVVVVVVVVVVVVVVVVVVVVVVVV
\section{Vanishing non-diagonal contribution}
\label{DSmoo}
%VVVVVVVVVVVVVVVVVVVVVVVVVVVVVVVVVVVVVVV

We consider a class of granular distributions close to being spatially uniform.
In the case of a cold plasma, such a distribution can be obtained by setting particles with a vanishing velocity over a cubic array.
For a multi-velocity distribution, we take a set of monokinetic beams where each beam is
a simple cubic array of particles whose elementary cube has its edges along the three orthogonal directions with coordinates $(x,y,z)$.
In the following, we are interested in a sequence of such sets of beams converging for $N$ large toward a spatially uniform Maxwellian distribution.
We now discretize this distribution into beams having the same number of particles.
The edge length for each beam is $L/n_{\mathrm{edge}}$ with $n_{\mathrm{edge}}$ an integer.
Therefore the number of particles in each beam is $n_{\mathrm{edge}}^3$.

In order to discretize the beam velocities, we first define in the space of velocities a sequence of $n_{\mathrm{shell}}$ shells
with increasing finite radii centered on the origin and containing each $N/n_{\mathrm{shell}}$ particles~;
the spacing between these radii is not constant, and increases like $1 / [ \| \bfv \|^2 f_0 (\| \bfv \|) ]$ as $\| \bfv \| \to \infty$.
This means the far tail of the Maxwellian is not described, but the range of described velocities increases with $n_{\mathrm{shell}}$.
We discretize the directions of the velocities into $n_{\mathrm{cone}}$ adjacent cones with a solid angle $4 \pi / n_{\mathrm{cone}}$.
These cones are chosen such that they fit into circular cones with a solid angle scaling like $n_{\mathrm{cone}}^{-1}$
(a particular instance of the bases of such cones can be obtained by tesselating the sphere as follows~:
choose a polar axis on the sphere, cut the latter with equidistant meridians,
and introduce a family of parallels defining equal annular areas on the sphere).
Therefore, the intersections of these cones with the shells define $n_{\mathrm{shell}} n_{\mathrm{cone}}$ velocity domains
with each $N/(n_{\mathrm{shell}} n_{\mathrm{cone}})$ particles,
whose extension in all directions vanishes when $n_{\mathrm{shell}}$ and $n_{\mathrm{cone}}$ go to infinity.
Each domain is related to one beam whose velocity is the average value of the velocity in the domain.
Then $N = n_{\mathrm{shell}} n_{\mathrm{cone}} n_{\mathrm{edge}}^3$.
The continuous distribution function is recovered in the limit
where $n_{\mathrm{shell}}$, $n_{\mathrm{cone}}$, and $ n_{\mathrm{edge}}$ all three go to infinity.
In order to describe particle distributions close to being spatially uniform,
we consider that the $\bfr_j(0)$'s and $\dot{\bfr}_j(0)$'s
are slightly shifted with respect to the $\bfr_{j0}$'s and $\bfv_j$'s.

We now focus on a given beam. It is convenient to consider the index $j$ of its particles as a vector $\bfp$
whose integer components run each from 1 to $n_{\mathrm{edge}}$.
The particles have the same value of $\bfv_{j}$.
Therefore, $\omega_{\bfm,j}$ and $\omega_{\bfn,j}$ take on a single value each in Eq.\ (\ref{phihat}),
and the summation over the $n_{\mathrm{edge}}^3$ particles of the beam of interest bears on
$\exp[\rmi \bfk_{\bfl} \cdot \bfr_{j0}]$ only, where $\bfl = \bfn-\bfm$.
Due to the periodicity of the $\bfr_{j0}$'s, the corresponding sum vanishes unless the three components of $\bfl$
are on the simple cubic lattice $(n_{\mathrm{edge}} {\mathbb{Z}})^3$ with mesh length $n_{\mathrm{edge}}$.
For the range of $k_{\bfm}$'s considered in section \ref{Mrdrv}, if $\pi b_{\mathrm{smooth}} n_{\mathrm{edge}}/L > 1$,
this occurs only for $\bfl = \mathbf{0}$~: the non-diagonal terms ($\bfn \neq \bfm$) in Eq.\ (\ref{phihatdiNdi}) vanish.
Then the inter-particle distance $d = L N^{-1/3} = L (n_{\mathrm{shell}} n_{\mathrm{cone}})^{-1/3} n_{\mathrm{edge}}^{-1} <  \pi b_{\mathrm{smooth}} (n_{\mathrm{shell}} n_{\mathrm{cone}})^{-1/3}$.
So, $d$ verifies condition (\ref{dllbsmoo}) when $n_{\mathrm{shell}}$ and $ n_{\mathrm{cone}}$
become large.

Conversely, if $d$ verifies condition (\ref{dllbsmoo}), let $n_{\mathrm{shell}} n_{\mathrm{cone}} \approx (b_{\mathrm{smooth}} / d)^{3\beta}$
for some $0 < \beta < 1/3$.
Recalling that $b_{\mathrm{smooth}} \ll \lambda_\rmD$, we find
  $n_{\mathrm{edge}} \approx N^{1/3} (b_{\mathrm{smooth}} / d)^{-\beta}
            \gg N^{(1-\beta)/3} (L / \lambda_\rmD)^{\beta}$,
which tends to infinity with $N$ independently from $L$ and $\lambda_\rmD$.
Thus the condition $d \ll b_{\mathrm{smooth}}$ is both sufficient and necessary for obtaining $n_{\mathrm{edge}} \gg 1$
and ensuring the vanishing of non-diagonal terms.

%TTTTTTTTTTTTTTTTTTTTTTTTTTTTTTTTTTTTTT
%============================================
% \begin{thebibliography}{99}
%============================================

% \section*{References}

\end{document}